# Core shift controls grain boundary energy scaling in Cu and Al

**Authors:** Xiaopu Zhang,[*] John J. Boland[*]

**Affiliations:**

Centre for Research on Adaptive Nanostructures and Nanodevices (CRANN), AMBER SFI Research Centre and School of Chemistry, Trinity College Dublin, Dublin 2, Ireland

*Correspondence to: xiaopuz@tcd.ie, jboland@tcd.ie

## Abstract

Grain boundary energies in different elements are correlated. The proportional scaling constants relating the energies of crystallography-equivalent boundaries in any two f.c.c. elements are nearly constant, with the notable exception of aluminum where these constants are known to vary significantly. However, the origins of the exceptional behavior of aluminum are not understood. Previously, we reported that for fcc metals there is a preference for boundaries to shift their tilt axis across the $(1\bar{1}0)$ plane towards [112] and to ultimately form low energy [112] core shifted boundaries (CSBs). By comparing grain boundary energies in copper and aluminum with different tilt axis in $(1\bar{1}0)$ plane, we now report the existence of a well-defined scaling behavior for the case of low angle boundaries. In contrast, the scaling constant for high angle boundaries is essentially fixed regardless of their tilt axis shift. This results in a gradual change in the scaling constants from low angle to high angle boundaries, which is responsible for the apparent exceptional scaling behavior found in aluminum. An analysis of structure evolution during core shifting points to the significance of boundary-core dissociation, a form of correlated relaxation of individual atoms at boundaries, in controlling the scaling of the boundary energies.

## Introduction

A fundamental goal of materials science is to understand the structure-property relationship, such as that between grain boundary structures and their energies. The anisotropy of boundary energies on boundary geometrical parameters plays an important role in driving microstructure evolution in the bulk and close to surfaces [1-12]. Recently, much attention has been paid to crystallography-equivalent (CE) boundaries, which have the same geometric thermodynamic variables – 5 macroscopic degrees of freedom. Atomic simulation shows that the energies of CE boundaries in different elemental metals show near proportional scaling [5-11].

Among the many proportional relationships between pairs of fcc metals, the scaling of boundary energies involving Al always show significant scatter from the proportional line and hence the behavior cannot be described by a well-defined scaling constant. While it is established that Al is different from other fcc metals, having a low elastic anisotropy such as $2C_{44}/(C_{11} - C_{12})$ and a high stacking fault energy $\gamma$, the role these differences play in the observed scaling behavior has never been addressed.

Here we studied the energy of Al boundaries with tilt axis from [552], through [111] and [112] to [113] in $(1\bar{1}0)$ plane, taking the corresponding Cu boundary energies as reference. We found that the scaling constant for low angle boundaries depends on the extent to which the tilt axes of these boundaries are shifted from [112], which we previously showed is always the lowest energy boundary for fcc metals regardless of the in-plane misorientation angle [13].

## Method

Boundary structures and their energies are determined by boundary geometries, microscopic shifts and individual atomic relaxations [1]. The effect of microscopic shifts was eliminated through structural searches that explored hundreds of different possible shifts. Hence, this allowed us to focus on the effect of individual atomic relaxations, and the influence of boundary microstructures on boundary energies.

We begin by considering symmetric [111] tilt boundaries with in-plane misorientation angle $\theta$ and $(1\bar{1}0)$ mean boundary plane. We calculated the boundary energies in the bulk for copper, aluminum, nickel, gold, and platinum at ten different angular shifts or inclinations of the composite rotation axis (CRA) away from [111] as the tilt axis shifts across the $(1\bar{1}0)$ boundary plane towards [112] and beyond [13]. We exploit here our new boundary analysis [13], which in contrast to that used earlier [3], is based on the orientation boundary cores and facilitate calculations at different tilts axes. We performed these calculations for 20 different in-plane angles [13] so that there are in total 200 CE boundaries considered in this work for each elemental metal. Calculations were performed with LAMMPS software, using molecular statics and the embedded-atom-method (EAM) interatomic potentials and the third-generation charge-optimized-many-body (COMB3) potential [14-19]. The boundary structures are visualized with OVITO [20].

## Results

### Pairwise comparisons

We show the pairwise comparison of CE grain boundary energies for Al and Cu in Fig. 1(a). Clearly there is a poor scaling relationship with many data points off the fitted proportional line, which is consistent with previous publications [7, 10]. The constant for these two groups of 200 CE grain boundaries is 1.71 close to the proportional constant of 1.783 derived from the data set calculated with the same EAM potentials by Tschopp et. al [10]. In contrast, the pairwise comparison of the same group of CE boundaries for Cu and Ni, that for Cu and Pt, shown in Fig. 1(b-c), and that for Cu and Au in Fig. S1, all show good proportional relationship as reported previously [7]. In Fig.1(d) we show the scaling of boundaries with [111] tilt axis, i.e., a core shift angle of zero. Clearly the scaling constant 1.63 for low angle boundaries (0 ~ 15°) and that 1.74 for all [111] boundaries (0 ~ 60°) are different.

### Scaling for different in-plane angles

To elucidate the dependence of the scaling behavior on the boundary structure at the atomic scale, we first classify all the data by the in-plane angle and show data points associated with different in-plane angles in Fig. 1(a-c) using different shapes and colors. Different points with the same shape and color correspond to the same in-plane boundary angle but with different composite rotation axes or core-shift angles in the $(1\bar{1}0)$ plane. We classify as subgroups boundaries with the same in-plane angle but different core-shift angles. A review of Fig. 1(a) shows that for the pairwise comparison of copper and aluminum each boundary subgroup shows different slopes with different finite non-zero intercepts. In contrast, Fig. 1(b-c) shows the pairwise comparison of copper and nickel and copper and platinum, which do not exhibit this type of behavior. Moreover, in Fig. 1(a) there is a systematic scatter for low angle boundaries with in-plane angle less than 15° and for median angle boundaries with in-plane angle between 15° and 32.20°. On the other hand, for high angle GBs with in-plane angles above 32.20°, the variation in the proportional scaling is considerably reduced. Figure 1(d) shows data points that exclude core shift in the $(1\bar{1}0)$ plane, so that all boundaries have [111] tilt axes, and from which the proportional constant for all and low angle boundaries are 1.74 and 1.63, respectively. This means that when exclusively low and even median angle [111] boundaries are considered, there is still a dependence of the scaling constant on the in-plane angle. A similar fit to the high angle boundary data yields a scaling constant of 1.80, as shown in Fig. 1(e). Remarkably, this scaling constant value (1.80) sets the upper limit for the combined low angle and median angle GB data sets, as shown in Fig. 1(f).

We now focus on the CE boundary energies for subgroups with in-plane angles of 3.89° and 16.43°. These boundaries are representative of the behavior of low angle boundaries and their scaling behavior are detailed in Fig. 2(a-b), respectively. For each in-plane angle, we consider 10 different core-shift angles within the $(1\bar{1}0)$ plane. Clearly the boundary energies vary according to the composite rotation axes and in both cases the energy reaches its minimum when the tilt axis lies along [112], which lies in the close packed $(11\bar{1})$ plane [13]. Since low angle boundaries are known to dissociate, this behavior reflects the effect of dislocation dissociation on boundary energies as the composite rotation axis shifts into and out of the close packed plane.

Figure 2(c) shows the slope and intercepts for all subgroups with low and median in-plane angles. Each subgroup exhibits a different slope and intercept, consistent with the earlier analysis by Tschopp et. al [10]. Low angle boundaries can be viewed as an array of dislocations, whose dissociation is sensitive to the core-shift angle within the ($1\bar{1}0$) plane. On the other hand, high angle boundaries do not dissociate and hence show no dependence on core-shift angle. In this scissor-shaped diagram there is a strong initial positive (negative) dependence of the slope (intercept) with increasing boundary angle, followed by an abrupt switch in this trend at $\theta = 19.65°$ and subsequently followed by an angle-independent behavior beyond $\theta = 38.21°$. In addition, we note the slope and intercept plots exhibit cusps at delimiting boundaries [4, 1] and [3, 1] in the structural unit model, reflecting the importance of atomic structures within the boundary cores [21, 22]. The slope of 1.80 found for the subgroup with even greater in-plane angles shown in Fig 2(c) is consistent with the fixed scaling constant for high angle boundaries found in Fig. 1(e) regardless of core shift angle.

**Core-shift angle dependent proportional constant**

Since the core-shift angle and the composite rotation axes are related to the jog density along boundary cores [13] and hence the degree to which dislocations can dissociate, we reclassify the data again into small groups according to their core-shift angle. Figure 3(a) shows data points for all 20 in-plane angles and 4 representative core-shift angles, $-19.47°$, $0°$, $19.47°$ and $29.50°$ that span the full range of core-shift angles considered in this work. The corresponding composite rotation axes in ($1\bar{1}0$) plane are [552], [111], [112] and [113], respectively. Axis [552] is the maximum shift away from [111] toward [110] that we have considered. Axis [112] is in the close-packed plane ($11\bar{1}$). Axis [113] is the maximum shift away from [111] toward [001]. For clarity, data points associated with the core-shift angles in-between are not shown. While Fig 3(a) shows evidence of scatter, the scaling behavior of GBs with the same core-shift is markedly better. To this end, Fig. 3(b) shows the data for the 7 low angles boundaries that comprised Fig 3(a) and from which the excellent scaling with core-shift angle is now evident. The proportional constants for core-shift angles $-19.47°$, $0°$, $19.47°$ and $29.50°$ are 1.77, 1.63, 1.31 and 1.54, respectively. The lowest energy boundaries are associated with a [112] composite rotation axis with a core-shift angle of $19.47°$ and these same boundaries exhibit the lowest possible scaling constant (1.31) Figure 3(c), shows the scaling constants for the low angle boundaries for each of the 10 core-shift angles considered in this study, with four data points taken from Fig. 3(b) highlighted. The dashed line corresponds to the scaling constant observed for high angle boundaries that is approached by low angles boundaries at increasingly negative core-shift angles. There is a monotonic variation of the scaling constant on either side of the $19.47°$ core-shift angle, which is associated with a [112] composite rotation axis lying in the close-packed plane ($11\bar{1}$).

To gain a better understanding of the dependence of the proportional scaling constant on the core-shift angle, we examined the atomic structure variation of CE boundaries in copper and aluminum with an in-plane angle of $3.89°$ at each of the 10 core-shift angles (Fig. 4). The atoms are selected by the centrosymmetric parameters and colored according to atomic energy. The grain boundaries are projected along their period vectors. Clearly, the copper dislocations are fully dissociated at the [112] composite rotation axis with two partials and stacking faults in-between. Atoms at the two partials have higher energy than those that comprise the stacking faults. In contrast, the aluminum dislocation at a [112] composite rotation axis is not dissociated, with the full dislocation core having the highest energy atoms at its center. This demonstrates that even without the confinement introduced by jogs that emerge at non-[112] composite rotation axes, stacking faults in aluminum are not dissociated [23]. In summary, the atoms in the copper boundary core are correlatively relaxed through the formation of extended stacking faults while the atoms in the aluminum boundary core are only locally relaxed. Even through the partial dislocation core in aluminum still can be identified [24], the two strongly-coupled partials are so close that the strain field and even the bonding distortion of two partials overlap, as shown in Fig. 4(b). The fact that the composite rotation axis lies in the close packed plane allows the formation of extended relaxed stacking fault structures in copper compared to the locally relaxed boundary core in aluminum and is responsible for the lowest possible proportional scaling constant.

When the composite rotation axis shifts away from [112] toward [552], the jog density in the copper boundary increases, breaking up the extended relaxation along the core, while the distance between the two partials decreases narrowing the width of the stacking fault ribbon. Gradually the dislocation become undissociated. In these cases, boundary-core atoms in both copper and aluminum are locally relaxed and thus the proportional scaling constant approaches the 1.802 value associated with high angle boundaries, where the stacking fault effects are diminished (or eliminated). A similar analysis also applies when the composite rotation axis shifts towards [113].

For the case of boundaries with median in-plane angles from 15° to 32.30°, as shown in Fig. 3(a), the dislocations are physically closer together so that their interaction increases in strength and the width of the splitting between partial becomes narrow so that a localized boundary core appears. Therefore, the relaxation mechanism of individual boundary atoms in copper and aluminum gradually become very similar, as shown in Fig. 1(e), so that the individual atomic relaxation is no longer important and only the crystallography plays a role, which yield a perfect proportional scaling. In this instance, the dependence of proportional constants on the core-shift angle gradually disappears and the overall scaling approaches that of high angle boundaries.

**Discussion**

In this work we have explored the possible origin of the anomalous scaling of Al grain boundary energies. The unusual behavior of aluminum is usually attributed to its high stacking fault energy and low elastic anisotropy [7]. However, nickel also has a high stacking fault energy, as shown in Tab. 1, and still exhibits well defined proportional scaling relationships with respect to other fcc metal as shown in Fig. 1(b) and in the literature [7, 23]. Hence the value of stacking fault energy cannot be the sole criterion. Considering the anisotropy of the Young's modulus, the shear modulus, and Poisson ratio, the quantification of the elastic anisotropy in terms of a single parameter is over simplified and unable to describe many physical properties [25], such as the differences in the proportional constants for different boundaries in one element.

Since the core-shift angle dependent scaling highlights the importance of the dissociation of dislocations, we explore the possibility that the dissociation width plays a role in scaling. The stacking fault width in isotropic materials is proportional to the ratio of the shear modulus to the stacking fault energy. Assuming the shear modulus can be described by $C_{44}$, closely related to the shear modulus in the dislocation model for the grain boundaries, we compared the ratio $C_{44}/\gamma$ for different elements and potentials and found that this ratio is possibly the most relevant for determining the scatter observed in the proportional scaling constant data [7, 26]. For example, the ratio for Al and Pt-COMB3 are 0.22 and 0.26, respectively. This suggests that the COMB3 potential for Pt, which is known to accurately describe the stacking fault energy in this metal, should show a similar behavior to that found in Al [19]. In contrast, the ratio for Cu, Ni and Pt EAM potentials are 1.7, 1.0 and 4.7, respectively. Hence for these materials, there are wider stacking faults in bulk and correspondingly less scatter in the pairwise scaling.

To demonstrate that a similar $C_{44}/\gamma$ ratio compared to Al yields similar scaling, we undertook a pairwise comparison of the CE boundary energies in copper using the EAM potential and platinum using the COMB3 potential. The full comparison is shown in Fig. S2. As was the case with Al, Fig. S2(a) shows that high angle grain boundaries exhibit better proportional scaling than low angle and median in-plane angle boundaries. In addition, Figure S2(c) shows that the low angle boundaries in platinum also exhibit core-shift angle dependent proportional scaling constants, the detailed analysis of which is shown for four different core-shift angles in Fig. S2(d). The core-shift angle dependent scaling constant in Fig. S2(e) exhibits the same kind of variation as that found for Al in Fig. 3(c). Clearly the detailed pairwise scaling of Cu and Pt using the COMB3 potential is strikingly similar to that found in Al, hence confirming the ratio $C_{44}/\gamma$ is a good predictor of the expected scaling behaviour.

Apart from local relaxation and dislocation dissociation, there are many other kinds of boundary relaxation phenomena such as structural unit formation, faceting, 9R-phase or stacking fault related boundary dissociation, and interface phase (complexion) transitions [27]. We believe that all these different types of relaxation should be reflected in the pairwise comparison [28]. Our calculation shows that [100] boundaries in copper and aluminum show similar kinds of core-shift angle dependent scaling constants but [110] boundaries show complex behavior of the proportional constant in the pairwise comparison, which is likely related to the stacking fault related boundary dissociation in copper [29, 30].

**Conclusion**

The scatter in the pairwise comparison of Cu and Al exhibits both an in-plane angle and core-shift angle dependence, which indicates that in addition to crystallography, different varieties of individual atom relaxation make an important contribution to grain boundary energy. The dislocation dissociation mechanism that correlates individual atomic relaxation in copper does not exist in aluminum, which is responsible for the large scatter in the pairwise comparison of CE boundaries, in particular low angle boundaries, where relaxation via dislocation dissociation is dominant.

J.J.B and X.Z. acknowledge support from Science Foundation Ireland grants (12/RC/2278 and 16/IA/4462) and thank Trinity Centre for High Performance for providing the computing resource.

**Figure 1 Comparison of grain boundary energies in Cu and Al.**
(a) Comparison of grain boundary energies calculated for copper and aluminum and proportional fit of all the data points. Data points are coloured by the in-plane angles. Multiple data point of the same colour and shape reflect the different core-shift angles. The slope of 1.71 is a best fit to all 200 CE boundaries. The 1.80 slope is the scaling for high angle boundaries. (b-c) Comparison of grain boundary energies calculated for nickel and platinum. (d) Comparison of grain boundary energies with composite rotation axis along [111] only. Proportional fit of low angle boundaries (black and blue data points) shown as a red line (slope 1.63). (e) Proportional fit of high angle boundaries energies with in-plane angles above 32.20. (f) Comparison of grain boundaries energies with in-plane angle less than **32.20°**.

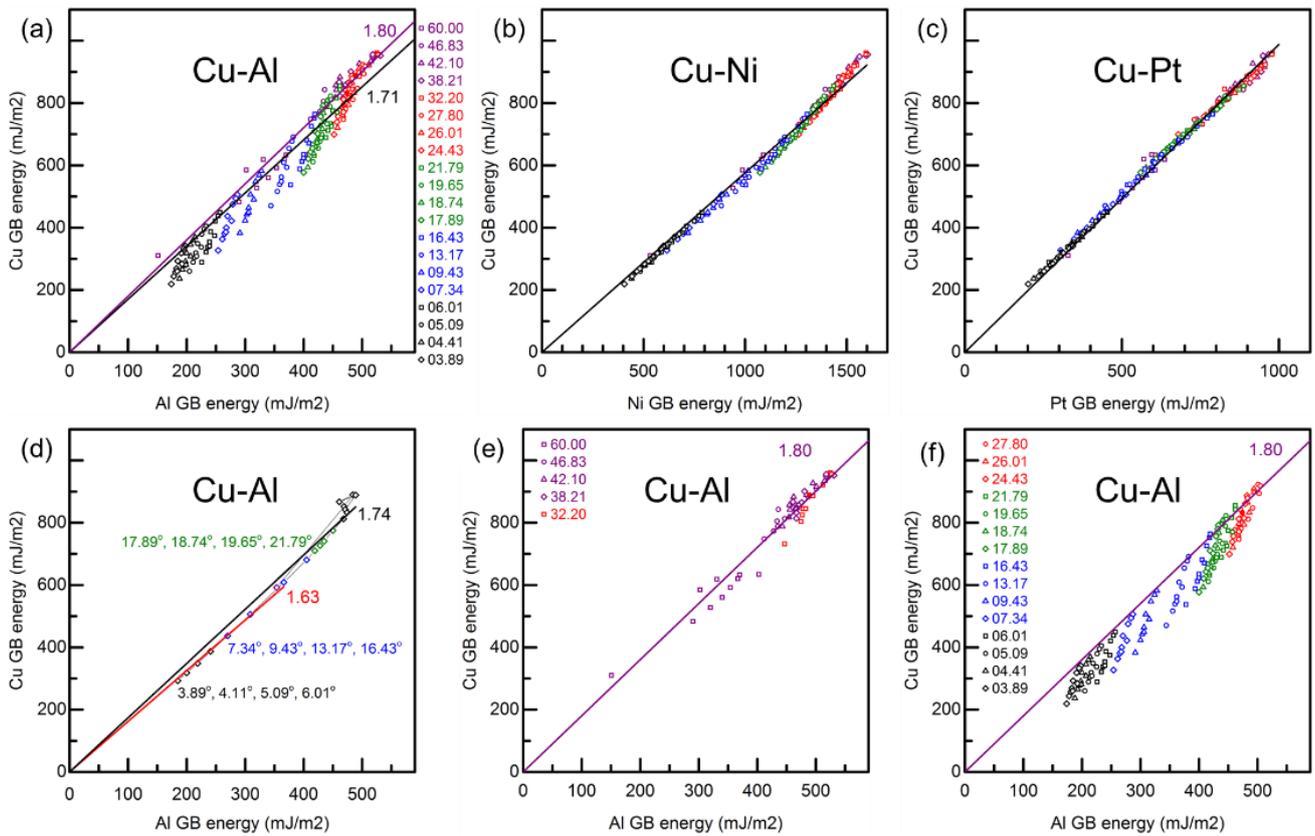

**Figure 2 Dependence of the linear fitting on in-plane angle.**
(a) Comparison of grain boundary energies with in-plane angle **3.89°** and different core-shift angles together with a linear fit (slope 4.09). (b) Comparison of grain boundary energies with in-plane angle **16.43°** and different core-shift angles together with a linear fit (slope 5.76). (c) the dependence of the slope and intercept on in-plane angles from the linear fitting of all in-plane angles up to 42.10º

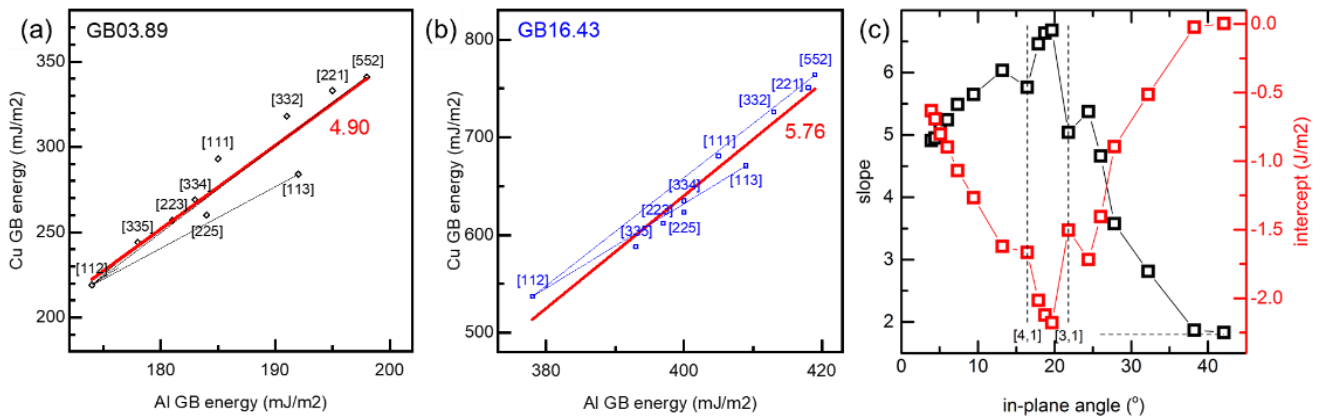

**Figure 3 Core shift dependent proportional constant.**
(a) Comparison of grain boundary energies for all in-plane angles with specified composite rotation axes or the core-shift angles referenced to [111]. (b) Proportional fitting of data points associated with seven low angle grain boundaries with composite rotation axis [552], [111], [112] and [113]. (c) Core-shift angle dependent proportional scaling constants as a function of core shift angle with the upper limit 1.80 (associated with large angle boundaries) shown as a dash line.

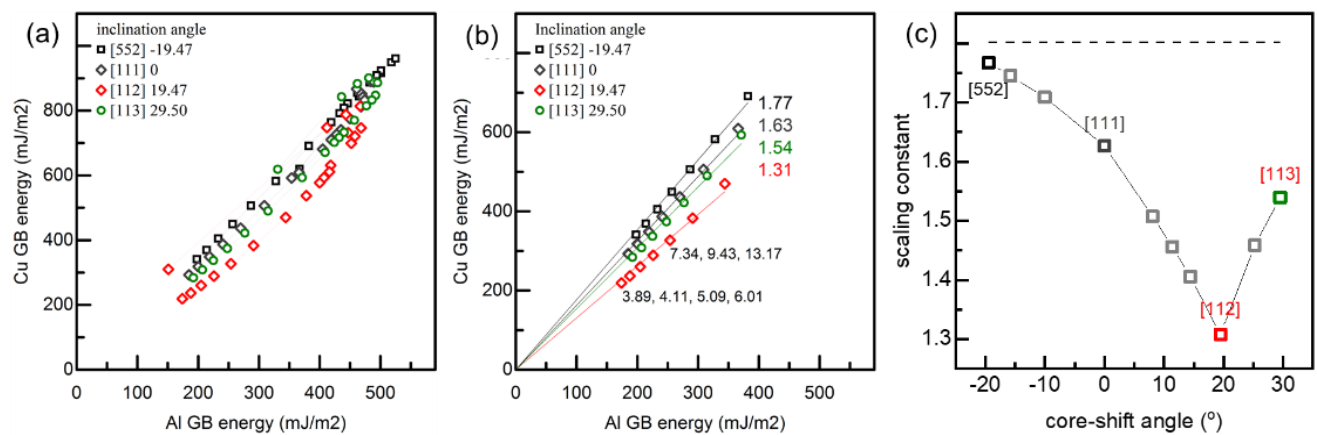

**Figure 4 Structural difference of GB3.89 in Cu and Al.**
(a) Variation of grain boundary structures in copper, projected along the boundary period vector, with different core-shift angles and the corresponding composite rotation axis. (b) Variation of grain boundary structures in aluminium. Non-fcc copper atoms are selected by centrosymmetric parameters greater than 0.39 and aluminium atoms greater than 0.49 [31]. All atoms are coloured by energy. The length of each stacking fault ribbon is 3 times of the length of the corresponding vector below graph (a).

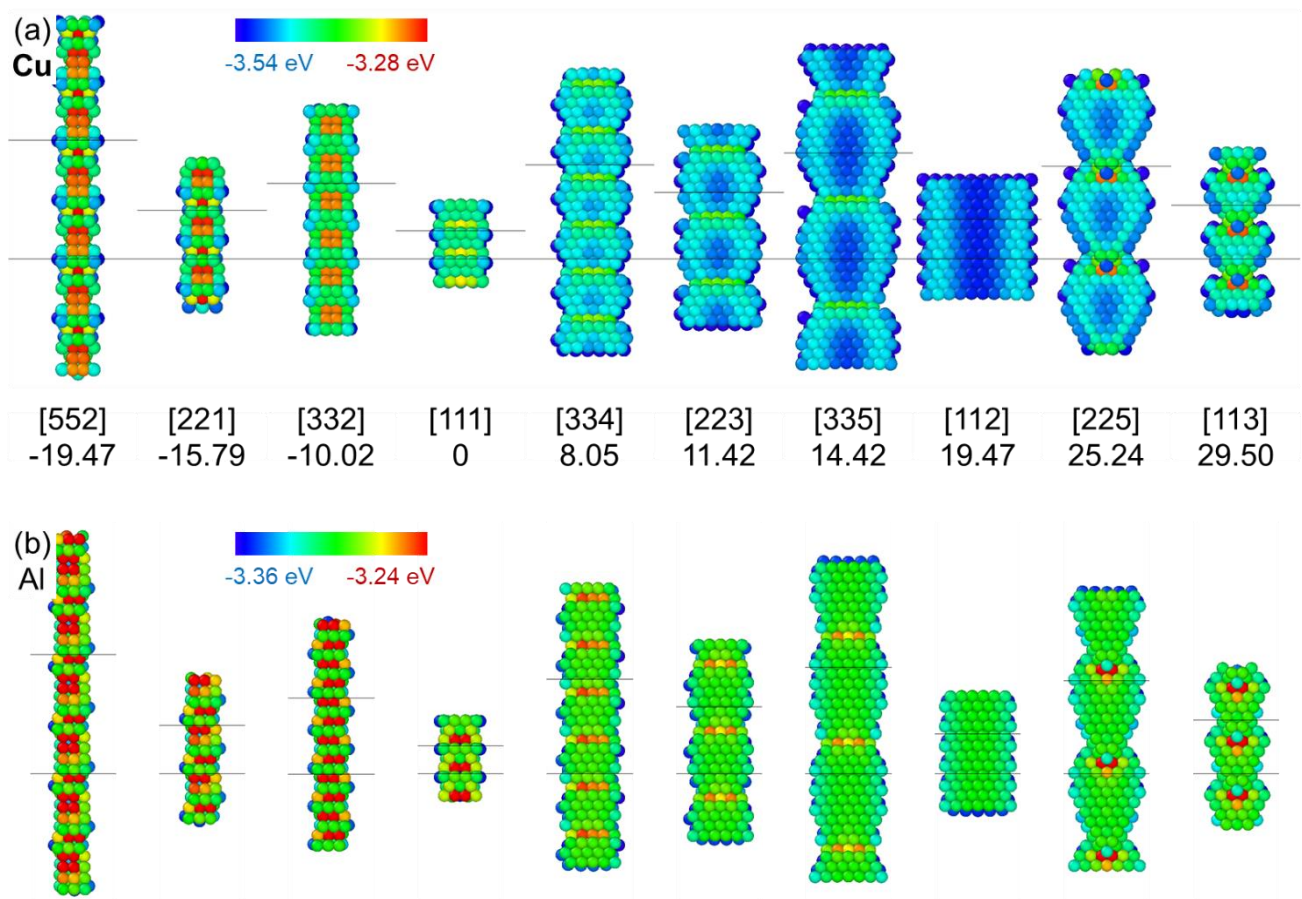

**Table 1 Calculated material properties from each potential for f. c. c. metal.**

|  | Au Foiles [17] | Cu Mishin [15] | Ni Mishin [16] | Al Mishin [16] | Pt Foiles [17] | Pt COMB3 [19] |
|---|---|---|---|---|---|---|
| Cohesive energy (eV) | -3.93 | -3.54 | -4.45 | -3.36 | -5.77 | -5.77 |
| Lattice constant (A) | 4.08 | 3.615 | 3.52 | 4.05 | 3.92 | 3.92 |
| Bulk modulus (GPa) | 166.9 | 138.4 | 181.2 | 80.0 | 283.1 | 228 |
| C11(GPa) | 183.2 | 169.9 | 247.9 | 113.8 | 303.1 | 283 |
| C12(GPa) | 158.8 | 122.6 | 147.8 | 61.6 | 273.1 | 198 |
| C44(GPa) | 44.7 | 76.2 | 124.8 | 31.6 | 68.3 | 82 |
| SF energy $\gamma$ | 4.7 | 44.4 | 125.2 | 145.5 | 14.61 | 321 |
| **C44/$\gamma$** | **9.5** | **1.7** | **1.0** | **0.22** | **4.7** | **0.26** |
| Zener anisotropy index | 3.7 | 3.2 | 2.5 | 1.2 | 4.6 | 1.9 |